\documentclass[aps,prb,preprint,superscriptaddress,endfloats]{revtex4}
\makeatletter
\def\@dotsep{4.5}
\makeatother

\usepackage{graphicx}
\usepackage{amsmath}	
\usepackage{bm}
\begin{document}


\title{Proximity effect between a dirty Fermi liquid and superfluid $^3$He}


\author{S. Higashitani}
\email[e-mail: ]{seiji@minerva.ias.hiroshima-u.ac.jp}
\affiliation{Graduate School of Integrated Arts and Sciences, Hiroshima
University, Kagamiyama 1-7-1, Higashi-Hiroshima 739-8521, Japan}

\author{Y. Nagato}
\affiliation{Information Media Center, Hiroshima University, Kagamiyama
1-4-2, Higashi-Hiroshima 739-8511, Japan}

\author{K. Nagai}
\affiliation{Graduate School of Integrated Arts and Sciences, Hiroshima
University, Kagamiyama 1-7-1, Higashi-Hiroshima 739-8521, Japan}

\date{\today}

\begin{abstract}
The proximity effect in superfluid $^3$He partly filled with high
porosity aerogel is discussed. This system can be regarded as a dirty
Fermi liquid/spin-triplet $p$-wave superfluid junction.  Our attention
is mainly paid to the case when the dirty layer is in the normal state
owing to the impurity pair-breaking effect by the aerogel.  We use the
quasiclassical Green's function to determine self-consistently the
spatial variations of the $p$-wave order parameter and the impurity
self-energy.  On the basis of the fully self-consistent calculation,
we analyze the spatial dependence of the pair function (anomalous
Green's function).
The spin-triplet pair function has in general even-frequency
odd-parity and odd-frequency even-parity components. We show that the
admixture of the even- and odd-frequency pairs occurs near the
aerogel/superfluid $^3$He-B interface.  Among those Cooper pairs, only
the odd-frequency $s$-wave pair can penetrate deep into the aerogel
layer.  As a result, the proximity-induced superfluidity in a thick
aerogel layer is dominated by the Cooper pair with the odd-frequency
$s$-wave symmetry.  We also analyze the local density of states and show
that it has a characteristic zero-energy peak reflecting the
existence
of the odd-frequency $s$-wave pair, in agreement with previous works
using the Usadel equation.
\end{abstract}


\maketitle


\section{Introduction\label{introduction}}

It is well known that the $p$-wave superfluid state of liquid $^3$He
near the container wall and the free surface is quite different from the
bulk
state.\cite{AmbegaokarPRA1974LGE,BuchholtzPRB1981IPS,NagatoJLTP1998RSE}
The quasiparticle scattering at the boundaries gives rise to substantial
pair-breaking and at the same time yields surface bound states.  Similar
phenomena take place also in unconventional superconductors.  The
existence of the surface bound states has been observed in a variety of
unconventional superconductors by tunneling spectroscopy experiments and
in superfluid $^3$He by transverse acoustic impedance
measurements.\cite{AokiPRL2005OSA,NagatoJLTP2007TTA}

Some anomalous superconducting properties due to 
the surface bound states have been predicted. 
The quasiparticle current carried by the surface bound states is so large 
that the Meissner current can be 
paramagnetic.\cite{HigashitaniJPSJ1997MPM}
Tanaka and Kashiwaya\cite{TanakaPRB2004ACT} have studied 
the effect on the charge transport in dirty normal metal/spin-triplet 
superconductor (DN/TS) junctions and found that 
the surface bound states cause an unusually small 
electric resistance of the junction. The Meissner effect in such a
dirty metal layer is also unusual.\cite{TanakaPRB2005AFP}

The DN/TS junctions have recently attracted much attention from the aspect
of a proximity-induced odd-frequency
pair.\cite{TanakaPRL2007TPE,TanakaPRB2007OFP,TanakaPRL2007AJE} In bulk
superconductors, the pair function (anomalous Green's function) is usually
even in the Matsubara frequency.  In general, however, it can be an odd
function; superconducting and superfluid states with this property are
referred to as odd-frequency pairing states.  Because of the Fermi
statistics, the spin and orbital symmetries of the odd-frequency pair are
classified into spin-singlet odd-parity and spin-triplet even-parity, in
contrast to the even-frequency pair.  Possibilities of the odd-frequency
pair in bulk superconductors have been discussed in the context of the
high-$T_c$ superconductors\cite{BalatskyPRB1992NCS,AbrahamsPRB1995POG} and
more recently for Ce-based compounds.\cite{FuseyaJPSJ2003ROF} In the DN
layer of the DN/TS junction, there is a possibility that an odd-frequency
spin-triplet superconductivity is generated by the proximity
effect.\cite{TanakaPRL2007TPE}

Superfluid $^3$He is the first material identified as a spin-triplet
pairing state. It is also the most well-understood unconventional
pairing
state and has played a role of a model system for understanding the
properties of newly discovered exotic superconductors.  This is mainly
because superfluid $^3$He is an extremely clean system. The effect of
impurities in superfluid $^3$He has also been studied intensively since the
discovery in 1995 of the superfluid transition of liquid $^3$He impregnated
in high porosity aerogel.\cite{PortoPRL1995SHA,SpraguePRL1995HES}

\begin{figure}
\includegraphics[width=10cm]{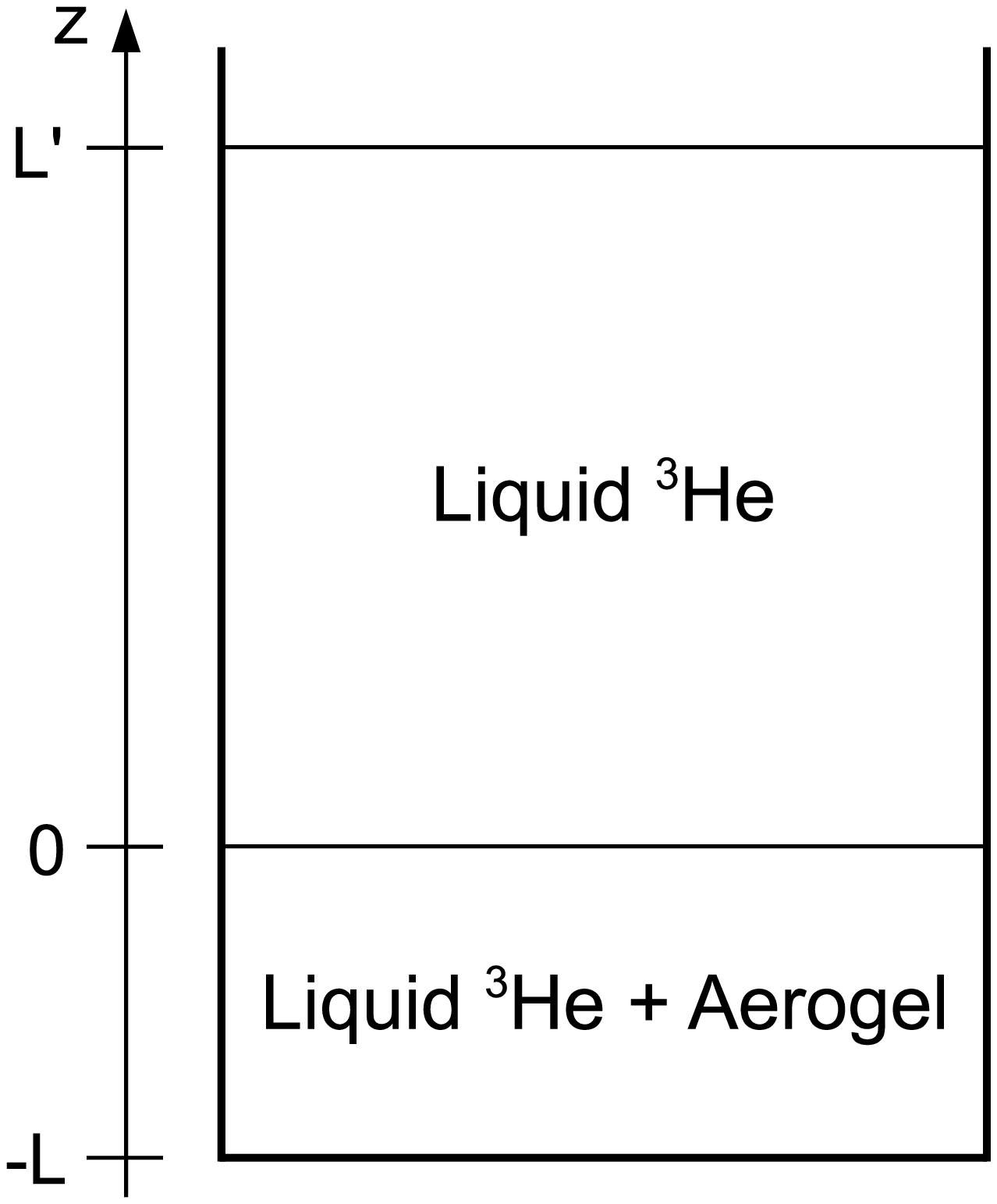} \caption{DN/TS system
consisting of liquid $^3$He and aerogel.  Liquid $^3$He in a container
occupies $-L < z < L'$ in which the aerogel is embedded in $-L < z < 0$.
When liquid $^3$He in the aerogel layer is in the normal state owing to
impurity effect and the pure liquid $^3$He layer is in a superfluid
state, this system is an ideal example of DN/TS junctions.
\label{system}}
\end{figure}

In this paper, we discuss the proximity effect in a DN/TS system
consisting of superfluid $^3$He and the aerogel, as depicted in Fig.\
\ref{system}.  Superfluid $^3$He is partly filled with the aerogel,
which is used to introduce short-range impurity potentials in liquid
$^3$He.  A measure of the impurity effect is a ratio $\xi_0/l$, where
$\xi_0$ is the coherence length and $l$ is the quasiparticle mean free
path.  A geometrical consideration\cite{ThunebergPRL1998MSH} for a
typical aerogel with 98 \% porosity leads to $l \sim$ 150 nm, which is
comparable to $\xi_0$ of superfluid $^3$He and therefore
a
strong impurity
effect is expected.  One of the remarkable characteristics of this
system is that the impurity effect can be controlled by pressure $P$:
the coherence length $\xi_0$ increases with decreasing pressure, so that
the system is relatively dirty at lower pressures.  As a consequence,
there exists a critical pressure $P_c$ below which the impurity effect
is so strong that superfluid transition does not occur down to zero
temperature.\cite{MatsumotoPRL1997QPT,RainerJLTP1998SPT,HigashitaniJLTP1999ISE}

One can estimate $P_c$ from Green's function theory for anisotropic pairing
states with randomly distributed
impurities.\cite{RainerJLTP1998SPT,HigashitaniJLTP1999ISE} The theory
predicts that the corresponding critical value of the ratio $\xi_0/l$ is
$e^{-\gamma}/2 \simeq 0.28$, where $\gamma = 0.577\cdots$ is Euler's
constant and the coherence length $\xi_0$ is defined by $\xi_0 = \hbar v_F
/ 2\pi k_B T_{c0}$ with $v_F$ the Fermi velocity and $T_{c0}$ the critical
temperature of pure liquid $^3$He.  Putting $l = 150$ nm gives a rough
estimate, $P_c \simeq$ 5 bar (at which $\xi_0 = 42$ nm), in reasonable
agreement with torsional oscillator experiments.\cite{MatsumotoPRL1997QPT}
The system as in Fig.\ 1 below $P_c$ is an ideal example of DN/TS
junctions.  Moreover, at high pressures above $P_c$, one can study
junctions with spin-triplet superfluids (TS'/TS system).

The purpose of this paper is to clarify the microscopic picture of the
proximity effect in the dirty normal Fermi liquid/superfluid $^3$He
system. A similar problem has been discussed by Kurki and
Thuneberg\cite{KurkiJLTP2007BCB} who calculated the self-consistent
$p$-wave order parameter near the interface using the Ginzburg-Landau
equation.  A theory applicable to low temperatures was proposed by
Tanaka {\it et al.}\cite{TanakaPRB2005TEP,TanakaPRB2004TCT} They derived
the boundary condition for the Usadel equation\cite{UsadelPRL1970GDE} at
the interface between the dirty normal metal and a clean unconventional
superconductor.  The Usadel equation treats
isotropic superfluidity
realized in dirty systems where non-$s$-wave pair functions are expected
to be suppressed by impurity scattering.  Near the interface, however,
any partial wave components of the pair function can in general coexist.
The Usadel equation cannot give information about such an interface
effect on a microscopic scale. In this paper, we give detailed numerical
results for the
spatial dependence
of the pair functions, obtained from the
quasiclassical Green's function theory valid for the whole temperature
range and for arbitrary values of the mean free path.  We also discuss
the local density of states in the dirty normal layer.

This paper is organized as follows. In Sec.\ \ref{formulation}, we
describe the theoretical model and the quasiclassical theory.  In Sec.\
\ref{numerical_results}, we present the numerical results for the
self-consistent $p$-wave order parameter, some partial wave components
of the pair function, and the local density of states. Both of the weak
scattering limit (Born approximation) and the strong one (unitarity
limit) are considered for the impurity effect. The final section is
devoted to conclusions.  In the rest of this paper, we use the units
$\hbar = k_B = 1$.

\section{Formulation\label{formulation}} 

We consider a dirty Fermi liquid/spin-triplet $p$-wave superfluid
junction as depicted in Fig.\ 1.  Liquid $^3$He in a container occupies
$-L < z < L'$ in which a region $-L < z < 0$ is filled with aerogel.
The container wall at $z=-L$ is simulated by a specular surface. The
system is assumed to have translational symmetry in the $x$ and $y$
directions (this implies that the random average is taken over impurity
positions).  The width $L'$ of the pure liquid $^3$He layer is assumed
to be much longer than the coherence length $\xi_0$ and we put $L'
\rightarrow \infty$.

Let us discuss the quasiclassical Green's function in the above system.
Because of the translational symmetry in the $x$ and $y$ directions, one
has only to consider one dimensional problem in the $z$ direction at a
given parallel momentum component $(p_x,p_y)$.  There exist two Fermi
momenta when the parallel momentum is fixed, namely, $(p_x,p_y,
+|p_{Fz}|)$ and $(p_x,p_y, -|p_{Fz}|)$, where $p_{Fz}$ is the $z$
component of the Fermi momentum.  Then, the quasiclassical Green's
function can be specified by three variables: the coordinate $z$, the
directional index $\alpha = {\rm sign}(p_{Fz})$, and the complex energy
variable $\epsilon$.

The quasiclassical Green's function, which we denote by 
$\hat g_\alpha(\epsilon,z)$, obeys\cite{SerenePhR1983QCA}
\begin{align}
\partial_z \hat g_\alpha(\epsilon,z) 
= [\hat h_\alpha(\epsilon,z),\ \hat g_\alpha(\epsilon,z)]
\label{Eeq}
\end{align}
with
\begin{eqnarray}
\hat h_\alpha(\epsilon,z) = \frac{\alpha i}{v_{Fz}}\left[
\begin{pmatrix}
\epsilon & \Delta_\alpha(z) \\
-\Delta_\alpha^\dagger(z) & -\epsilon \\
\end{pmatrix}
- \hat\sigma^{\rm imp}(\epsilon,z)\right],
\end{eqnarray}
where $v_{Fz}$ is the $z$ component of the Fermi velocity, 
$\Delta_\alpha$ is the $2\times 2$ matrix of the $p$-wave order parameter, 
and $\hat\sigma^{\rm imp}$ is the impurity self-energy.
The quasiclassical Green's function $\hat g_\alpha$ is normalized as
$\hat g^2_\alpha = -1$.

We assume superfluid $^3$He to be in the B phase.
In this case, the order parameter can be expressed as
\begin{align}
&\Delta_\alpha(z) = 
\begin{pmatrix}
-\Delta_0(z)\sin\theta\, e^{-i\phi} & \alpha \Delta_1(z)\cos\theta \\
\alpha \Delta_1(z)\cos\theta & \Delta_0(z)\sin\theta\, e^{i\phi} \\
\end{pmatrix},
\label{eq:b_phase_order_parameter}
\end{align}
where $\theta$ and $\phi$ are the polar and azimuthal angles, respectively.
Note that in our notation the sign of $p_{Fz} = p_F\cos\theta$
is specified by $\alpha$ and therefore the polar angle $\theta$ is
in the range $0 < \theta < \pi/2$.  For $z \rightarrow \infty$, the
spatially dependent order parameters $\Delta_0(z)$ and $\Delta_1(z)$ tend
to the same constant value, $\Delta_B$, corresponding to the energy gap in
the bulk B phase (we take $\Delta_B$ to be real).

We describe the impurity effect within the self-consistent $t$-matrix
approximation\cite{BuchholtzPRB1981IPS,HigashitaniJLTP1999ISE,HigashitaniPRB2005MTS}
for randomly distributed delta-function potentials.  This formulation
allows one to discuss both of the weak scattering regime (Born
approximation) and the strong one (unitarity limit).  In the two limits,
the impurity self-energies are given by
\begin{align}
\hat\sigma^{\rm imp} = 
\begin{cases}
\displaystyle
-\frac{1}{2\tau}\hat G & \mbox{(Born limit)}\\
\\
\displaystyle
\frac{1}{2\tau}\hat G^{-1} & \mbox{(Unitarity limit)}\\
\end{cases}
\label{sigma}
\end{align}
with 
\begin{align}
\hat G =  \frac{1}{2}\sum_{\alpha=\pm}\langle \hat g_\alpha \rangle
\end{align}
and $\tau = l/v_F$ the mean free time. Here,
\begin{align}
\langle \cdots \rangle = 
\int_0^{\pi/2} \sin\theta\, 
d\theta \int_0^{2\pi} \frac{d\phi}{2\pi}\,(\cdots).
\end{align}
In the system under consideration, the impurity scattering rate 
$1/\tau$ changes discontinuously at 
the interface: it is zero in the region of pure superfluid $^3$He 
($z > 0$) and finite otherwise ($-L < z < 0$).

Equation (\ref{Eeq}) is supplemented by the following boundary conditions. 
In the asymptotic region ($z \rightarrow \infty$), 
the quasiclassical Green's function takes the 
bulk form
\begin{align}
\hat g_\alpha(\epsilon,z) \xrightarrow{z \rightarrow \infty} 
\frac{1}{\sqrt{\Delta_B^2 - \epsilon^2}}
\begin{pmatrix}
\epsilon & \Delta_\alpha(\infty) \\
-\Delta_\alpha^\dagger(\infty) & -\epsilon \\
\end{pmatrix}.
\end{align}
At the interface ($z = 0$), $\hat g_\alpha$ is continuously connected. 
At the wall ($z = -L$), it satisfies the specular surface boundary 
condition
\begin{align}
\hat g_+(\epsilon, -L) = \hat g_-(\epsilon, -L).
\end{align} 

The above set of equations enables one to determine the quasiclassical
Green's function $\hat g_\alpha$ in the present system.  As shown by Nagato
{\it et al.}\cite{NagatoJLTP1993TAR,NagatoJLTP1996TRS} and Higashitani and
Nagai,\cite{HigashitaniJPSJ1995MEN} the calculation of $\hat g_\alpha$ can
be reduced to solving simple and numerically stable 
differential equations of the Riccati type. 
They treated the cases of $2\times 2$ matrix Green's functions.  A similar
problem for $4\times 4$ matrix cases was discussed by
Eschrig.\cite{EschrigPRB2000DFN}
Following them,
we write $\hat
g_\alpha$ as
\begin{align}
\hat g_\alpha = 
\begin{pmatrix}
[1 + {\cal D}_\alpha \tilde{\cal D}_\alpha]^{-1} & 0 \\
0 & [1 + \tilde{\cal D}_\alpha {\cal D}_\alpha]^{-1} \\
\end{pmatrix}
\begin{pmatrix}
i[1-{\cal D}_\alpha\tilde{\cal D}_\alpha] & 2{\cal D}_\alpha \\
-2\tilde{\cal D}_\alpha & -i[1-\tilde{\cal D}_\alpha{\cal D}_\alpha] \\
\end{pmatrix}
\end{align}
in which ${\cal D}_\alpha$ and $\tilde{\cal D}_\alpha$ 
are $2\times 2$ matrices obeying
the Riccati type differential equations
\begin{align}
&\alpha v_{Fz} \partial_z {\cal D}_\alpha 
= i(a_\alpha{\cal D}_\alpha + {\cal D}_\alpha\tilde{a}_\alpha)
 + b_\alpha - {\cal D}_\alpha\tilde{b}_\alpha{\cal D}_\alpha, 
\label{Ricone}\\
&\alpha v_{Fz} \partial_z \tilde{\cal D}_\alpha 
= -i(\tilde{a}_\alpha\tilde{\cal D}_\alpha + \tilde{\cal D}_\alpha a_\alpha)
 - \tilde{b}_\alpha + \tilde{\cal D}_\alpha{b}_\alpha\tilde{\cal D}_\alpha.
\label{Rictwo}
\end{align}
Here, we have put
\begin{align}
\hat h_\alpha = \frac{\alpha i}{v_{Fz}}
\begin{pmatrix}
a_\alpha & b_\alpha \\
-\tilde{b}_\alpha & -\tilde{a}_\alpha \\
\end{pmatrix}.
\end{align}

We can further simplify the calculation 
by taking into account the matrix structure of the order parameter 
in the B phase. One can write $\Delta_\alpha$ as 
\begin{align}
\Delta_\alpha = V \sigma_1[(\alpha \Delta_\perp +
i\Delta_\parallel)\sigma_+
+(\alpha \Delta_\perp - i \Delta_\parallel)\sigma_-]V,
\label{delta}
\end{align}
where
\begin{align}
&\Delta_\parallel = \Delta_0 \sin\theta,\\
&\Delta_\perp = \Delta_1 \cos\theta,\\
&V = \exp\left(-\frac{i}{2}\phi\,\sigma_3\right),\\
&\sigma_\pm = \frac{1}{2}(1 \pm \sigma_2),
\end{align}
and $\sigma_i$'s ($i = 1,2,3$) are the Pauli matrices.
Note that $\sigma_\pm$ are projection operators satisfying 
\begin{align}
\sigma_\pm^2 = \sigma_\pm,\quad
\sigma_+\sigma_- = \sigma_-\sigma_+ = 0,\quad
\sigma_+ + \sigma_- = 1.
\end{align}
Equation (\ref{delta}) suggests the following parameterization:
\begin{align}
&{\cal D}_\alpha = V\sigma_1({\cal F}_\alpha \sigma_+ + 
\tilde{\cal F}_\alpha\sigma_-)V,
\label{dparaone}\\
&\tilde{\cal D}_\alpha = -V^\dagger({\cal F}_{-\alpha} \sigma_+ + 
\tilde{\cal F}_{-\alpha}\sigma_-)\sigma_1 V^\dagger,
\label{dparatwo}
\end{align}
where ${\cal F}_\alpha$ and $\tilde{\cal F}_\alpha$ are scalar functions.
Using Eqs.\ (\ref{dparaone}) and (\ref{dparatwo}), 
the angle-averaged Green's function $\hat G$ is found to have the form
\begin{align}
&\hat G = 
\begin{pmatrix}
iG & F\sigma_1 \\
F\sigma_1 & -iG \\
\end{pmatrix}
\label{ghat}
\end{align}
with
\begin{align}
&G = \frac{1}{2}\left\langle
\frac{1 + \lambda}{1-\lambda}  
+\frac{1 + \tilde\lambda}{1-\tilde\lambda}\right\rangle,
\label{g_swave}\\
&F = \frac{1}{2}\left\langle
\frac{{\cal F}_+ + {\cal F}_-}{1-\lambda}  
+\frac{\tilde{\cal F}_+ + \tilde{\cal F}_-}{1-\tilde\lambda}
\right\rangle,
\label{f_swave}
\end{align}
where 
$\lambda = {\cal F}_+{\cal F}_-$ and
$\tilde\lambda = \tilde{\cal F}_+\tilde{\cal F}_-$.
The function ${\cal F}_\alpha$ obeys 
\begin{align}
\alpha v_{Fz}\partial_z {\cal F}_\alpha 
= 2i h {\cal F}_\alpha + f_\alpha + f_{-\alpha}{\cal F}_\alpha^2,
\end{align}
where
\begin{align}
&h = \epsilon + \frac{i}{2\tau}G,\\
&{f}_\alpha  = \alpha\Delta_\perp + i \Delta_\parallel + \frac{1}{2\tau}F.
\end{align}
The boundary conditions are
\begin{itemize}
\item[(a)] $z \rightarrow \infty$:\ 
$\displaystyle{
{\cal F}_- = \lim_{z \rightarrow \infty}
\frac{if_-}{h + i \sqrt{-f_+f_- -h^2}}
}$.
\item[(b)] $z=0$: ${\cal F}_\alpha$ is continuous.
\item[(c)] $z=-L$: ${\cal F}_+ = {\cal F}_-$.
\end{itemize}
The function $\tilde{\cal F}_\alpha$ obeys the same set of 
equations as ${\cal F}_\alpha$ but with $\Delta_\parallel$ replaced 
by $-\Delta_\parallel$.
Thus, the $4 \times 4$ matrix equation for $\hat g_\alpha$ is reduced to 
the four scalar ones for ${\cal F}_\pm$ and $\tilde{\cal F}_\pm$.

It is useful to note that 
${\cal F}_\alpha(\epsilon,z)$ and $\tilde{\cal
F}_\alpha(\epsilon,z)$ on the imaginary axis ($\epsilon=iE$) of the complex
$\epsilon$ plane satisfy the symmetry relation
\begin{align}
&{\cal F}_\alpha^*(iE,z) = \tilde{\cal F}_\alpha(iE,z),
\label{fmatone}\\
&{\cal F}_\alpha(-iE,z) = 1/\tilde{\cal F}_\alpha(iE,z).
\label{fmattwo}
\end{align}
Equation (\ref{fmatone}) allows one to construct the Matsubara Green's
function only from ${\cal F}_\alpha$.  It follows also from Eq.\
(\ref{fmatone}) that the angle averaged Green's functions $G$ and $F$ at
$\epsilon=iE$ are real quantities [see Eqs.\ (\ref{g_swave}) and
(\ref{f_swave})].  Equation (\ref{fmattwo}) is associated with the parity
of Green's function in the Matsubara frequency $\epsilon=i\epsilon_n = i\pi
T(2n+1)$.  For example, one can show that the spin-triplet $s$-wave pair
function $F$ is odd in $\epsilon_n$, as expected.

Finally we discuss briefly how we can treat the case when the B-phase
order parameter in the asymptotic region takes a more general form
$\Delta_\alpha(z \rightarrow \infty) = \Delta_B R_{\mu j}({\bf n},
\vartheta) \hat p_j \sigma_\mu i \sigma_2$, where $\hat p_j =
p_j/p^{}_F$ and $R_{\mu j}({\bf n}, \vartheta)$ is a rotation matrix
around an axis ${\bf n}$ by an angle $\vartheta$. Our analysis of the
quasiclassical Green's function relies on a particular choice of the
order parameter [see Eq.\ (\ref{eq:b_phase_order_parameter})],
corresponding to $R_{\mu j} = \delta_{\mu j}$. It is easy to generalize
our formulation to the case of arbitrary $R_{\mu j}$. Since $R_{\mu
j}\sigma_\mu = U \sigma_j U^\dagger$ with $U = \exp(-i
\bm{\sigma}\cdot{\bf n} \vartheta/2)$, the quasiclassical Green's
function for arbitrary $R_{\mu j}$ can be obtained from the one for
$R_{\mu j} = \delta_{\mu j}$ by a spin-space rotation described by
$U$. This means that the difference between the quasiclassical Green's
functions for different choice of $R_{\mu j}$ is only the matrix
structure in spin space. For example, the (1,2)-component (in
particle-hole space) of the angle-averaged Green's function $\hat G$ for
$R_{\mu j} = \delta_{\mu j}$ has the form $F \sigma_1 = F \sigma_3 i
\sigma_2$ [see Eq.\ (\ref{ghat})] and is transformed by the rotation to
$F U \sigma_3 U^\dagger i \sigma_2 = F R_{\mu 3} \sigma_\mu i \sigma_2$.
In the present study of the proximity effect such a difference in the
spin structure is not important and all the numerical results presented
below are the same regardless of the choice of $R_{\mu j}$.

\section{Numerical results\label{numerical_results}}

In this section, we show some numerical results relevant to the proximity
effect.  We discuss, in separate subsections, the self-consistent $p$-wave
order parameter, the odd-frequency $s$-wave pair function, non-$s$-wave
pair functions, and the local density of states. The numerical results
given there are those obtained within the Born approximation. The effect of
the strong scattering shall be discussed in the final subsection.  All the
numerical calculations are performed at a temperature $T = 0.2T_{c0}$
($T_{c0}$ is the critical temperature of pure liquid $^3$He), which is low
enough for the $p$-wave order parameter in the bulk superfluid $^3$He to be
well developed.

\subsection{Self-consistent $p$-wave order parameter} 

The $p$-wave order parameter must be determined self-consistently from 
the gap equation with the pairing interaction 
$v_{\hat p,\hat p'} = -3g_1{\hat p}\cdot{\hat p}'$, where ${\hat p}$ is the 
unit vector specifying the direction of the Fermi momentum. 
The gap equation can be written in terms of ${\cal F}_\alpha$ as 
\begin{align}
&\frac{\Delta_0(z)}{g_1N(0)} = 3\pi T \sum_{\epsilon_n > 0}^{\epsilon_c}
\left\langle \sin\theta\, {\rm Im}\,A^+(i\epsilon_n,z)
\right\rangle,\\
&\frac{\Delta_1(z)}{g_1N(0)} = 6 \pi T \sum_{\epsilon_n > 0}^{\epsilon_c}
\left\langle \cos\theta\, {\rm Re}\,
A^-(i\epsilon_n,z)\right\rangle,\\
&A^\pm(i\epsilon_n,z) = \frac{{\cal F}_+(i\epsilon_n,z)
\pm {\cal F}_-(i\epsilon_n,z)}
{1-\lambda(i\epsilon_n,z)},
\end{align}
where $N(0)$ is the density of states at the Fermi level and 
$\epsilon_c$ is a cutoff energy.

We have calculated the self-consistent order parameter by solving the above
equations iteratively. In the numerical calculations, the coupling constant
is evaluated from\cite{KieselmannPRB1987SCC}
\begin{align}
\frac{1}{g_1N(0)} = \ln\frac{T}{T_{c0}} 
+ 2\pi T \sum_{\epsilon_n>0}^{\epsilon_c}\frac{1}{\epsilon_n}
\end{align}
and the Matsubara summation is cut off at $\epsilon_c = 10 \pi T_{c0}$.

\begin{figure}
\includegraphics[width=10cm]{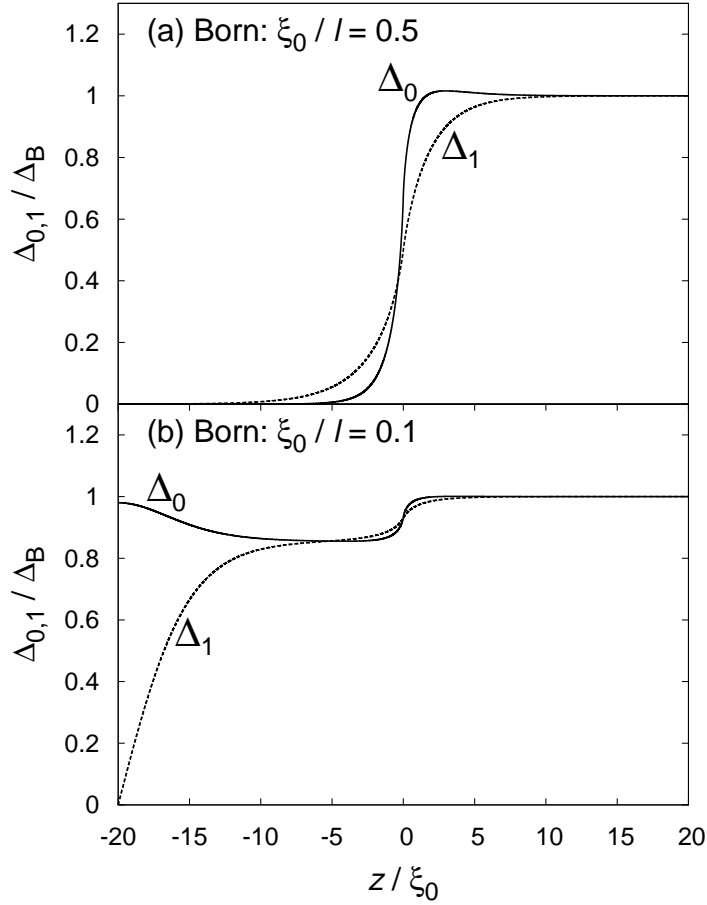}%
\caption{
Spatial variation of the self-consistent $p$-wave order parameter
in the Born limit with (a) $\xi_0/l = 0.5$ and (b) $\xi_0/l = 0.1$.
The vertical axis is $\Delta_{0,1}$ scaled by the bulk gap 
$\Delta_B$ of superfluid $^3$He-B.
The $^3$He-aerogel layer occupies $-20 < z/\xi_0 < 0$ and is in contact
with bulk superfluid $^3$He-B at $z = 0$.
In the aerogel layer far from the interface,
impurity scattering destroys the $p$-wave Cooper pair 
for $\xi_0/l = 0.5$, but does not for $\xi_0/l = 0.1$.
\label{scop}}
\end{figure}

Typical results of the spatial dependence of $\Delta_0(z)$ and
$\Delta_1(z)$ are shown in Fig.\ \ref{scop}.
We take the width $L$ of the aerogel layer to be $20\xi_0$.
The impurity effect by the aerogel is characterized by  
the parameter $\xi_0/l$ and is evaluated within the Born approximation.
Numerical results for two cases with $\xi_0/l =0.1, 0.5$ are shown.  
When $l = 150$ nm (see Sec.\ \ref{introduction}), 
the parameters $\xi_0/l = 0.1$ and $\xi_0/l = 0.5$
correspond to pressures $\sim 34$ bar (above the critical pressure $P_c$) 
and $\sim 0$ bar (below $P_c$), respectively. 

For $\xi_0/l = 0.5$, the impurity effect is strong enough to 
destroy bulk $p$-wave superfluidity of $^3$He.
In this case, the $p$-wave order parameters in the aerogel layer 
exhibit exponential decay typical for the proximity effect.

In the relatively clean case with $\xi_0/l = 0.1$, the $p$-wave order
parameters survive even in the region far from the interface.  Thus, the
surface effect characteristic of the B phase is expected near the wall
(at $z/\xi_0=-20$).\cite{BuchholtzPRB1981IPS,NagatoJLTP1998RSE} The
strong suppression of $\Delta_1$ and the enhancement of $\Delta_0$ near
the wall are due to the surface effect. Near the interface, on the other
hand, both of $\Delta_0$ and $\Delta_1$ are suppressed by the impurity
effect.

\subsection{Odd-frequency $s$-wave pair function\label{nresult_odd_freq}}

In this subsection, we discuss the odd-frequency spin-triplet 
$s$-wave pair 
corresponding to the function $F$ 
[see Eqs.\ (\ref{ghat}) and (\ref{f_swave})].  
Here, we are interested in
$F$ in the normal liquid 
$^3$He-aerogel layer. 

First of all, we mention
a
qualitative picture expected from the
dirty-limit theory.  The DN layer is characterized by three parameters:
the diffusion constant $D=v_Fl/3$, the Matsubara frequency $\epsilon_n$,
and the layer width $L$, as can be seen from the Usadel
equation.\cite{UsadelPRL1970GDE} From dimensional analysis, we have a
characteristic length scale $\xi_d(\epsilon_n) = \sqrt{D/\epsilon_n}$
and a characteristic energy scale $E_{\rm Th} = D/L^2$ (the Thouless
energy).
The length $\xi_d(\epsilon_n)$ is essentially equivalent to the
temperature-dependent coherence length $\xi_N = \sqrt{D/2\pi T}$ in the
DN layer. As has been shown in the study of the proximity effect in
metals,\cite{deGennes,Deutscher} $\xi_N$ gives the length scale of the
exponential decay of the proximity-induced pair function in the dirty
limit.
The frequency-dependent coherence length $\xi_d(\epsilon_n)$
characterizes the spatial variation of $F$ in the dirty limit.  At high
frequencies $\epsilon_n \gg E_{\rm Th}$, $\xi_d(\epsilon_n)$ is much
shorter than $L$.  This means that $F$ is exponentially small near the
layer end.  On the other hand, at low frequencies $\epsilon_n \ll E_{\rm
Th}$, $F$ extends over the whole layer.  For $\epsilon_n \rightarrow 0$,
in particular, $\xi_d(\epsilon_n)$ diverges, so that a long range
proximity effect is expected.

\begin{figure}
\includegraphics[width=15cm]{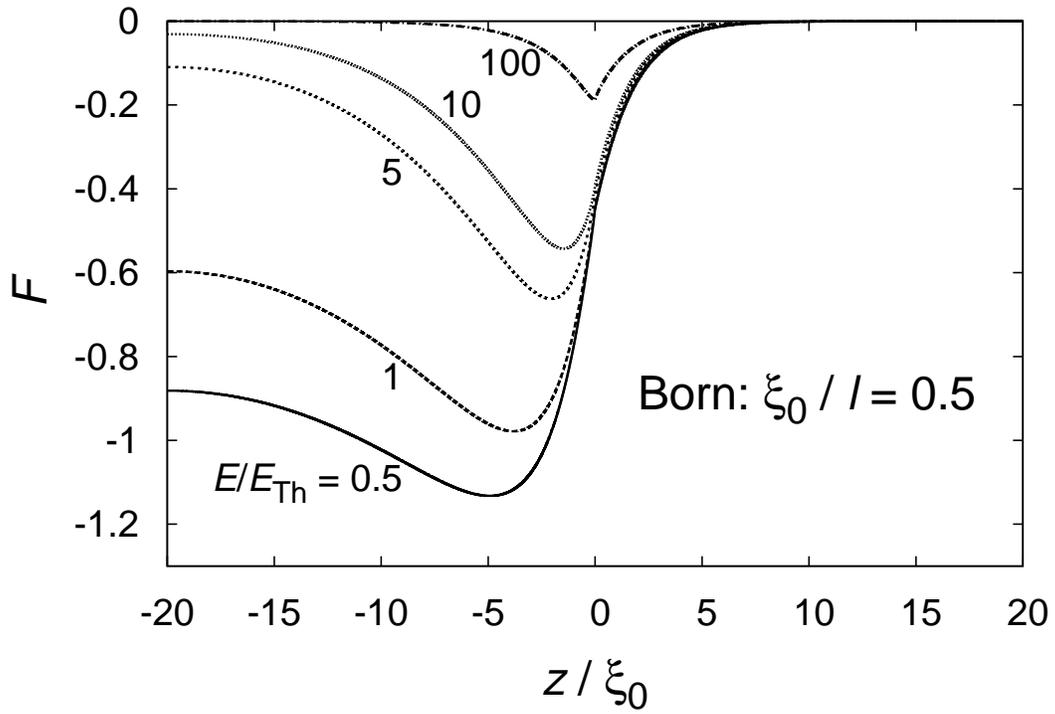}%
\caption{Spatial dependence of the odd-frequency spin-triplet 
$s$-wave pair function $F(iE,z)$ at various values of $E/E_{\rm Th}$ 
in the same DN/TS system as in Fig.\ \ref{scop}(a).
\label{odd_freq_pair}}
\end{figure}

In Fig.\ \ref{odd_freq_pair}, our numerical results for $F(\epsilon,z)$
are shown for various values of imaginary frequencies ($\epsilon = iE$)
scaled by $E_{\rm Th}$. The impurity effect is evaluated at $\xi_0/l =
0.5$ in the Born limit.  The numerical results support the above
qualitative picture. At a high frequency $E/E_{\rm Th} = 100$, $F$ is
localized near the interface.  With decreasing frequency, the
penetration distance of $F$ increases. The magnitude of $F$ also
increases with decreasing frequency.

We note here that the odd-frequency $s$-wave pair function $F$ does not
appear when the superfluid layer ($z > 0$) does not have the
perpendicular component $\Delta_\perp$ of the $p$-wave order parameter.
This can be analytically shown in the following way. Putting
$\Delta_\perp = 0$ and assuming that $F = 0$, one can readily find a
relation $\tilde{\cal F}_\alpha = -{\cal F}_\alpha$.  Then, 
the upper-right $2\times 2$ submatrix of $\hat g_\alpha$ (which we denote 
by $\hat g^{12}_\alpha$) takes the form
\begin{align}
\hat g_\alpha^{12} = \frac{2i{\cal F}_\alpha}{1-\lambda}
\begin{pmatrix}
e^{-i\phi} & 0 \\
0 & -e^{i\phi} \\
\end{pmatrix}.
\end{align} 
This form of $\hat g_\alpha^{12}$ is consistent with the assumption of
$F=0$.  Thus, when $\Delta_\perp =0$, the pair function has only the
$p$-wave components parallel to the interface.

\subsection{Proximity effect of non-$s$-wave pair functions\label{non-s-wave}}

There can coexist any partial wave components of the pair function 
near the interface that breaks translational symmetry. 
Here, we give numerical results of the spatial dependence of 
the $p$-wave and $d$-wave components.

The $p$-wave components have even-frequency spin-triplet symmetry 
and summing them up over the Matsubara frequencies gives 
the $p$-wave order parameter. One can define three $p$-wave components 
$F_1^{x,y,z}$ as 
\begin{align}
&\frac{1}{2}\sum_{\alpha=\pm}
\left\langle \hat p_x\,\hat g_\alpha^{12}\right\rangle
= F_1^{x} \sigma_1 i \sigma_2,\\
&\frac{1}{2}\sum_{\alpha=\pm}
\left\langle \hat p_y\, \hat g_\alpha^{12}\right\rangle
= F_1^{y} \sigma_2 i \sigma_2,\\
&\frac{1}{2}\sum_{\alpha=\pm}
\left\langle \alpha\hat p_z\,\hat g_\alpha^{12}\right\rangle
= F_1^{z} \sigma_3 i \sigma_2.
\end{align}
It is obvious from the symmetry of the system that 
\begin{align}
F_1^x = F_1^y.
\end{align}

The $d$-wave components are odd-frequency spin-triplet pairs, as well as 
the $s$-wave one. As an example of the $d$-wave components, we consider 
\begin{align}
\frac{1}{2}\sum_{\alpha=\pm}
\left\langle \frac{1}{2}(3\hat p_z^2 -1)
\hat g_\alpha^{12}\right\rangle
= F_2 \sigma_3 i \sigma_2.
\end{align}

\begin{figure}
\includegraphics[width=15cm]{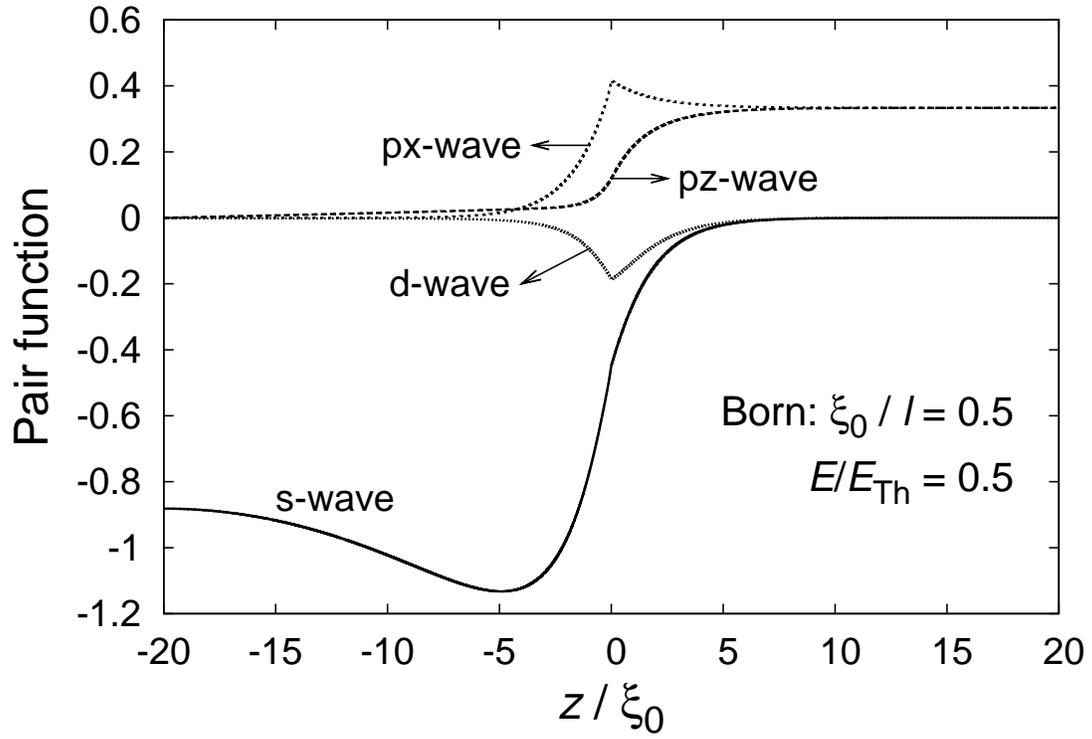}%
\caption{Spatial dependence of some partial wave components of the 
pair function at an imaginary frequency $\epsilon=iE$ with 
$E/E_{\rm Th} = 0.5$ 
in the same DN/TS system as in Fig.\ \ref{scop}(a).
\label{non_s_wave}}
\end{figure}

In Fig.\ \ref{non_s_wave}, we show the spatial dependence of the
$p$-wave and $d$-wave pair functions for $\xi_0/l = 0.5$ in the Born
limit.  The corresponding result
for
the $s$-wave pair function is also
shown for comparison. All the components coexist around the interface.
In the DN layer except near the interface, the non-$s$-wave components
are much smaller than the $s$-wave one and the proximity-induced
superfluidity is dominated by the odd-frequency $s$-wave pair.

\subsection{Local density of states}

\begin{figure}
\includegraphics[width=15cm]{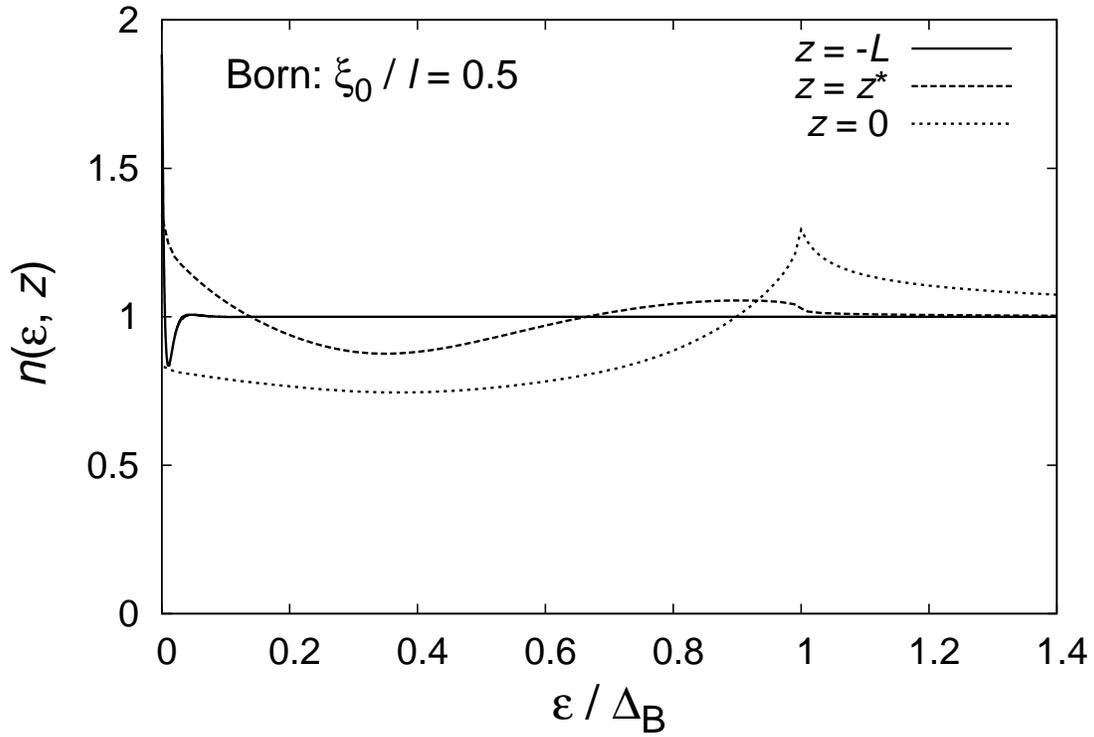} 
\caption{LDOS in the same DN/TS system as
in Fig.\ \ref{scop}(a).  Three lines are those at the DN layer end
(solid line), at the interface (dotted line), and at $z =
z^*$ (dashed line), a position near the interface in the DN layer, so
defined that $\Delta_1(z^*)/\Delta_1(0) = 1/2$.
\label{ldosn}}
\end{figure}

We now discuss the local density of states (LDOS),
\begin{align}
n(\epsilon, z) ={\rm Re}\,G(\epsilon + i0,z).
\end{align}
Figure \ref{ldosn} shows LDOS in the Born limit with $\xi_0/l = 0.5$.
LDOS at the interface (dotted line) has a peak at $\epsilon/\Delta_B =
1$ above which it decreases with increasing $\epsilon$ and tends to
unity (corresponding to the normal-state value) for $\epsilon
\rightarrow \infty$.  The finite LDOS at low energies $\epsilon/\Delta_B
< 1$ is due to the existence of bound states decaying into the pure
superfluid layer.  At the layer end (solid line), on the other hand,
LDOS has a sharp peak at zero energy and is unity in almost all the
other energies.  At a position near the interface (dashed line), LDOS
shows an intermediate behavior between the two cases.

\begin{figure}
\includegraphics[width=15cm]{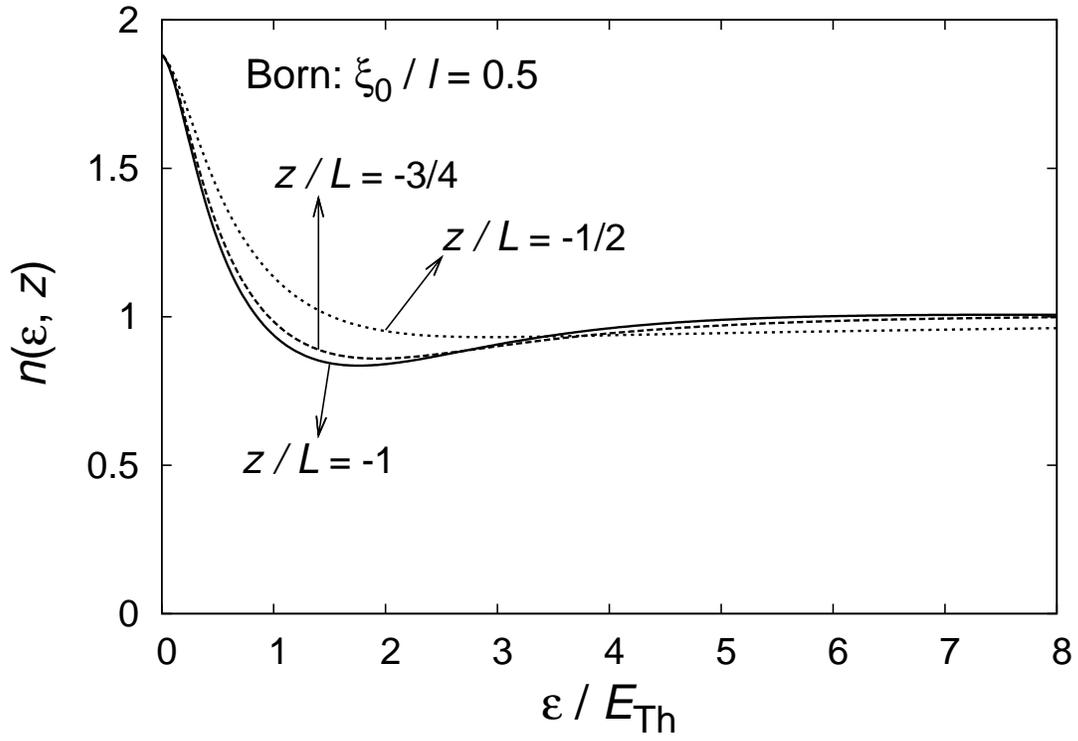}%
\caption{LDOS as a function of $\epsilon/E_{\rm Th}$.
Numerical results at three positions $z/L = -1,-3/4,-1/2$ 
in the DN layer are shown for the same DN/TS system
as in Fig.\ \ref{scop}(a).
\label{zeroenergy}}
\end{figure}

The width of the zero-energy peak in LDOS at the layer end is of order
the Thouless energy $E_{\rm Th}$. In the present system, $E_{\rm Th}$ is
much smaller than $\Delta_B$ ($E_{\rm Th}/\Delta_B \sim 6\times
10^{-3}$).  The peak structure around $\epsilon=0$ is magnified in Fig.\
\ref{zeroenergy}. Here, we plot LDOS at $z/L=-1, -3/4, -1/2$ as a
function of $\epsilon/E_{\rm Th}$.  It is, at first glance, surprising
that a sizable peak appears in LDOS near the layer end where the
$p$-wave order parameter almost vanishes [see Fig.\ \ref{scop}(a)].  It
should be noted, however, that at low energies below $\epsilon \sim
E_{\rm Th}$ the $s$-wave pair function has a finite amplitude even near
the layer end though non-$s$-wave pair functions are vanishingly small
there (see Fig.\ \ref{non_s_wave}).  The zero-energy peak structure
below $\epsilon \sim E_{\rm Th}$ in LDOS in the DN layer is due to the
existence of the odd-frequency $s$-wave pair.

\subsection{Effect of strong impurity scattering}

All the above numerical results are calculated within the Born
approximation.  Here, we discuss the strong scattering effect as
described by the impurity self-energy in the unitarity limit.

As is well known, the impurity effect in anisotropic superfluid states
is quite different between the weak and strong scattering regimes.  In
our theory such a difference originates from the anisotropy (momentum
dependence) of the quasiclassical Green's function $\hat g_\alpha$.
Note that, when $\hat g_\alpha$ is isotropic, the angle average of $\hat
g_\alpha$, namely, $\hat G$ has a property $\hat G^2 = -1$ similar to
the normalization condition $\hat g_\alpha^2=-1$.  Then, the impurity
self-energy in the unitarity limit has the same form as that in the Born
limit [see Eq.\ (\ref{sigma})].

\begin{figure}
\includegraphics[width=15cm]{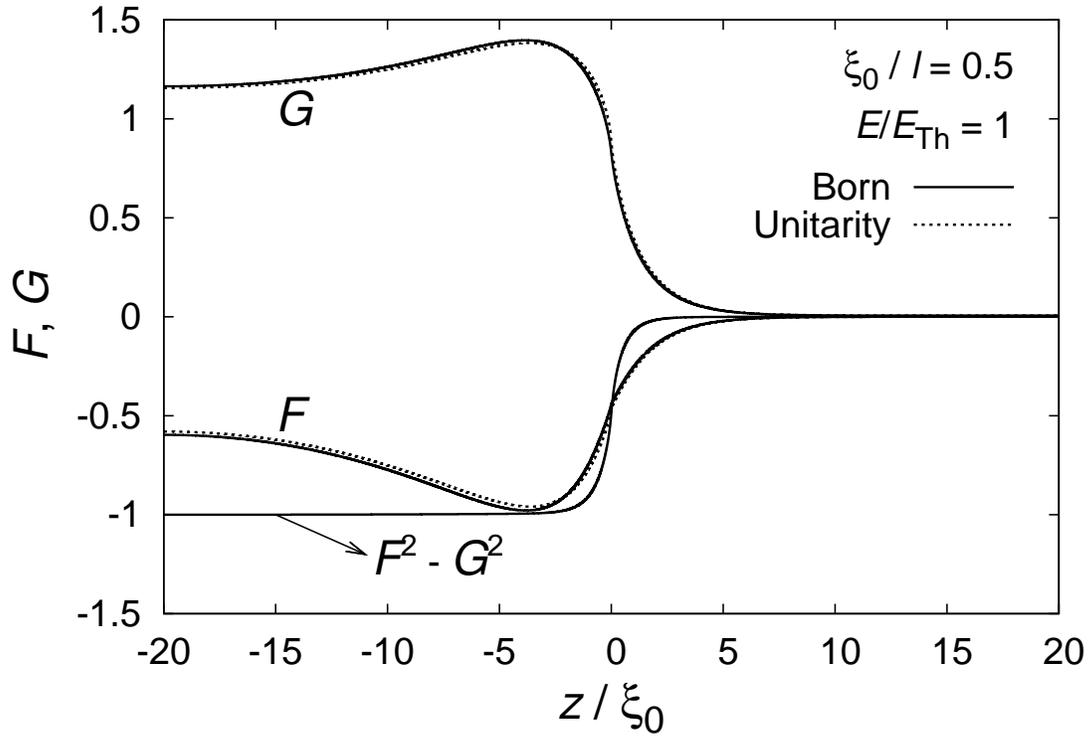}%
\caption{Comparison between the Born-limit (solid lines) and 
the unitarity-limit (dotted lines) results for the spatial dependence 
of the matrix elements $G$ and $F$ of the angle averaged Green's function 
$\hat G(iE,z)$ at $E/E_{\rm Th}=1$ for $\xi_0/l = 0.5$. 
The results for $G$ and $F$ in the unitarity limit are almost the same 
as those in the Born limit. 
The spatial dependence of $\hat G^2 = F^2 - G^2$ in the Born limit 
is plotted for reference (see text).
\label{s-wave-comp}}
\end{figure}

The above observation implies that the impurity effects in the DN layer
are insensitive to scattering regime because of isotropization by
impurity scattering. To demonstrate that, we plot in Fig.\
\ref{s-wave-comp} the matrix elements $G$ and $F$ of $\hat G$ for
$\xi_0/l=0.5$ as a function of $z/\xi_0$ in both of the Born and
unitarity limits.  The difference between the two scattering limits is
quite small and $\hat G^2 = F^2 - G^2 = -1$ is well satisfied in the DN
layer except near the interface.  We have compared the other Born-limit
numerical results presented previous subsections for $\xi_0/l = 0.5$
with the corresponding results in the unitarity limit and found also no
significant difference.

\begin{figure}
\includegraphics[width=12cm]{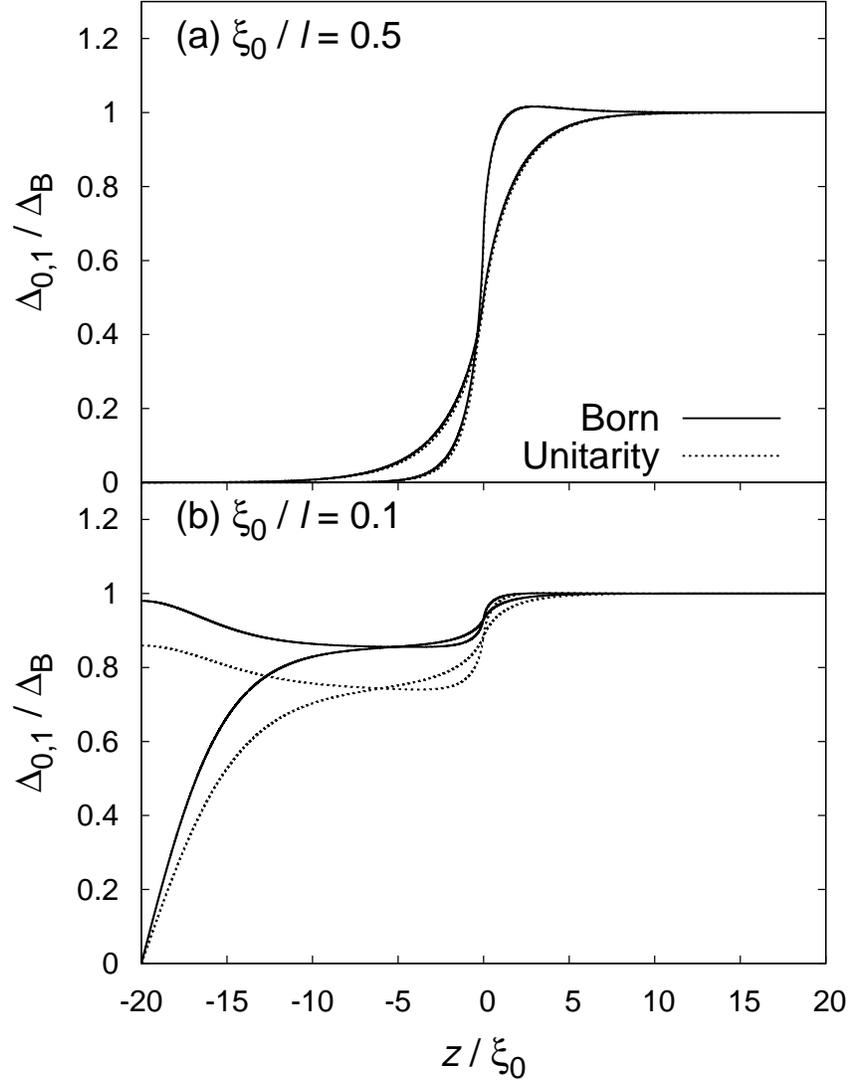}%
\caption{Comparison between the Born-limit (solid lines) and 
the unitarity-limit (dotted lines) results for the spatial dependence of 
the self-consistent $p$-wave order parameter.
The numerical data in the Born limit are the same as those 
in Fig.\ \ref{scop}. 
For $\xi_0/l = 0.5$, the results in the two limits 
are almost the same.
\label{p-wave-comp}}
\end{figure}

On the other hand, when $\xi_0/l$ is sufficiently small and liquid
$^3$He in the aerogel is in the $p$-wave superfluid state, substantial
differences
can occur.  As an example, we show in Fig.\ \ref{p-wave-comp}
the spatial dependence of the $p$-wave order parameter. The
unitarity-limit scattering gives rise to stronger pair-breaking. This
effect is clearly seen for
$\xi_0/l = 0.1$
but is quite small for
$\xi_0/l = 0.5$.

\section{Conclusions}

Using the quasiclassical Green's function theory, we have discussed the
proximity effect in the liquid $^3$He-aerogel composite in contact with
bulk superfluid $^3$He-B.  This system is an ideal model system for
studying the proximity effect in dirty normal Fermi liquid/spin-triplet
superfluid junctions when the impurity pair-breaking effect by the
aerogel is strong enough to destroy the superfluidity of liquid $^3$He.
Such a situation occurs at low pressures below the critical
pressure\cite{MatsumotoPRL1997QPT,RainerJLTP1998SPT,HigashitaniJLTP1999ISE}
or at temperatures higher than the reduced critical temperature of
liquid $^3$He in the aerogel.  We have proposed a convenient Green's
function parameterization that reduces the $4\times 4$ matrix equation
for superfluid $^3$He-B to four scalar differential equations of the
Riccati type.  We used this method to calculate self-consistently the
spatial variations of the $p$-wave order parameter and the impurity
self-energy.  To clarify the microscopic structure of the
proximity-induced superfluidity, we have analyzed in detail the
frequency and momentum dependent pair function and the local density of
states in the aerogel layer.

Although the order parameter in the above system is
purely $p$-wave,
any partial wave components of the pair function can coexist owing to
broken translational symmetry.  We have shown that such an admixture of
the Cooper pairs, in fact, occurs around the aerogel/superfluid $^3$He-B
interface.  Among those, only the $s$-wave pair can penetrate deep into
the aerogel layer.  This is because non-$s$-wave pairs are destroyed by
impurity scattering in the interior of the aerogel layer.  As is
expected qualitatively from the Usadel equation and was confirmed by our
fully self-consistent numerical calculations, the $s$-wave pair function
extends over the whole aerogel layer when the frequency is of order or
less than the Thouless energy $E_{\rm Th}$.  The $s$-wave pair in the
present system has odd-frequency symmetry and therefore the
proximity-induced superfluidity in a thick aerogel layer is an example
of the odd-frequency pairing states.

We have found that the local density of states in the aerogel layer has
a zero-energy peak of width of order $E_{\rm Th}$, in agreement with
previous theoretical
works\cite{TanakaPRB2004ACT,TanakaPRB2005AFP,TanakaPRB2005TEP,TanakaPRL2007TPE}
using the Usadel equation.  The aerogel layer considered here has such a
thick width that the $p$-wave order parameter almost vanishes near the
layer end.  Nonetheless, a sizable zero-energy peak appears in the local
density of states near the layer end.  This is a manifestation of the
long range proximity
effect
of the $s$-wave pair function at frequencies below 
$\epsilon \sim E_{\rm Th}$.

One of the remaining problems which we have not addressed in the present
paper is what characteristics in observables arise from the
odd-frequency $s$-wave paring.  Further studies in this direction are in
progress.

\begin{acknowledgments}
This work is supported in part by a Grant-in-Aid for Scientific Research on
Priority Areas (No.\ 17071009) from the Ministry of Education, Culture,
Sports, Science and Technology of Japan.  We would like to thank Y. Tanaka
and Y. Asano for useful discussions.
\end{acknowledgments}




\newpage

\end{document}